\documentclass[fleqn,usenatbib]{mnras}

\usepackage{newtxtext,newtxmath}

\usepackage[T1]{fontenc}
\usepackage{ae,aecompl}

\usepackage{graphicx}	
\usepackage{amsmath}	
\usepackage{amssymb}	
\usepackage{booktabs}
\usepackage{longtable}
\usepackage{subfig}
\usepackage{bm}
\usepackage{threeparttable}
\usepackage[export]{adjustbox}

\newcommand{\ha}{$\mathrm{H\alpha}$}	

\title[Ram pressure candidates in Coma]{Ram pressure stripping candidates in the Coma Cluster: Evidence for enhanced star formation}

\author[I. D. Roberts et al.]{
Ian D. Roberts\thanks{E-mail: roberid@mcmaster.ca},
Laura C. Parker
\\
Department of Physics and Astronomy, McMaster University, Hamilton ON L8S 4M1, Canada\\
}

\date{Accepted XXX. Received YYY; in original form ZZZ}

\pubyear{2019}

\begin{document}
\label{firstpage}
\pagerange{\pageref{firstpage}--\pageref{lastpage}}
\maketitle

\begin{abstract}
The Coma cluster is the nearest massive ($M \gtrsim 10^{15}\,\mathrm{M_\odot}$) galaxy cluster, making it an excellent laboratory to probe the influence of the cluster environment on galaxy star formation.  Here, we present a sample of 41 galaxies with disturbed morphologies consistent with ram pressure stripping. These galaxies are identified visually using high-quality, multi-band imaging from the Canada-France-Hawaii telescope covering $\sim\!9\,\mathrm{deg^2}$ of the Coma cluster.  These ``stripping candidates'' are clear outliers in common quantitative morphological measures, such as concentration-asymmetry and Gini-$M_{20}$, confirming their disturbed nature.  Based on the orientations of observed asymmetries, as well as the galaxy positions in projected phase-space, these candidates are consistent with galaxies being stripped shortly after infall onto the Coma cluster.  Finally, the stripping candidates show enhanced star formation rates, both relative to ``normal'' star-forming Coma galaxies and isolated galaxies in the field.  Ram pressure is likely driving an enhancement in star formation during the stripping phase, prior to quenching.  On the whole, ram pressure stripping appears to be ubiquitous across all regions of the Coma cluster.
\end{abstract}




\section{Introduction}
\label{sec:intro}

Galaxy properties in the local Universe display a persistent bimodality.  The vast majority of local galaxies are either: (1) galaxies which have blue colours, are gas-rich, are actively star-forming, and have late-type morphologies; or (2) galaxies which have red colours, are gas-poor, show little-to-no ongoing star formation, and have early-type morphologies.  Conventional thinking suggests that these two populations of galaxies represent different points along an evolutionary sequence, with blue, star-forming galaxies (``blue cloud'') evolving into red, passive galaxies (``red sequence'').  Galaxies which exhibit properties intermediate to the blue cloud and red sequence (known as the ``green valley'') are often considered transition galaxies currently experiencing star formation quenching \citep[e.g.][]{salim2014}.  The fact that the green valley is sparsely populated relative to the blue cloud and red sequence, suggests that this transition (quenching) is likely quite rapid.  This model works on average, but there are many nuances and exceptions which complicate this simple picture.  Some galaxies may transition from the red sequence onto the blue cloud, as opposed to the other direction, due to star formation being rejuvenated \citep[e.g.][]{clemens2009,chauke2019}.  Additionally, star formation rates for passive galaxies are often upper limits, which introduces substantial uncertainties in star formation properties for galaxies off main sequence.
\par 
Understanding the physical mechanisms which are driving this star formation quenching requires large, diverse samples of galaxies which span wide ranges in stellar mass and local environment.  For example, the fraction of quiescent galaxies depends strongly on stellar mass such that higher mass galaxies are increasingly quiescent whereas low-mass galaxies are far more likely to be star-forming \citep[e.g.][]{peng2010, geha2012, wetzel2012}.  The origin of this trend with stellar mass is often ascribed to processes internal to the galaxy such as feedback from supernovae or AGN as well as the high virial temperature of massive galaxy halos impeding the cooling of gas \citep[e.g.][]{dekel2006,schawinski2009}.  These trends with stellar mass are well established, however even at fixed stellar mass, galaxy star formation shows a clear trend with local environment \citep[e.g.][]{wetzel2012}.  For example, quiescent fractions scale with halo mass such that galaxies residing in massive clusters are preferentially quenched compared to field galaxies of the same mass.  Even within individual clusters a clear environmental dependence is present, as galaxy populations in the central cluster region are dominated by quiescent galaxies relative to the cluster outskirts \citep[e.g.][]{postman2005, blanton2007, prescott2011, rasmussen2012, fasano2015, haines2015}.  The balance between internally and externally driven quenching is also a clear function of galaxy mass, with low-mass galaxies ($\log\,M_\mathrm{star} \lesssim 10 - 10.5\,\mathrm{M_\odot}$) being significantly more quenched by environment, whereas quenching higher mass galaxies is more strongly associated with internal processes \citep[e.g.][]{haines2006, bamford2009, peng2010}.
\par
Galaxy clusters represent the most-massive virialized objects in the local Universe, making them the ideal place to probe environmentally-driven galaxy evolution.  These extreme environments are capable of rapidly shutting down star formation \citep[e.g.][]{wetzel2012,haines2015,brown2017,roberts2019} in member galaxies.  Many physical mechanisms have been proposed for quenching star formation in galaxy clusters, which can be roughly divided into two classes: (1) interactions between galaxies and the hot, X-ray emitting intracluster medium (ICM) permeating the cluster; and (2) dynamical interactions between cluster galaxies or between galaxies and the cluster halo.  Examples belonging to the first category include: ram pressure stripping \citep[e.g.][]{gunn1972, quilis2000}, viscous stripping \citep{nulsen1982}, and starvation/strangulation \citep[e.g.][]{larson1980, peng2015}; whereas examples of dynamical interactions include: mergers \citep[e.g.][]{mihos1994a, mihos1994b}, harassment \citep[e.g.][]{moore1996}, and tidal interactions \citep[e.g.][]{mayer2006, chung2007}.  While all of these mechanisms are capable of affecting cluster galaxies, the key question is which are the primary mechanisms driving quenching and does the dominant mechanism change with halo mass?
\par
Recently, ram pressure stripping and starvation have been favoured for driving quenching in groups and clusters \citep[e.g.][]{muzzin2014, peng2015, fillingham2015, wetzel2015, brown2017, foltz2018, vanderburg2018, roberts2019}.  Ram pressure stripping involves the direct removal of cold-gas from galactic discs as galaxies traverse the ICM at high speeds.  Signatures of ram pressure stripping include tails of stripped gas or stars trailing behind galaxies in clusters \citep{mcpartland2015,poggianti2017}, as well as star forming discs which appear truncated from the outside-in \citep{schaefer2017,finn2018,schaefer2019}.  On the other hand, starvation is the removal of the gas reservoir for future star formation.  This can occur due to the high virial temperature of the cluster preventing hot halo gas from cooling and condensing onto galactic discs, or by the removal of this halo gas from galaxies through stripping.  Evidence for starvation can be inferred from galaxy metallicities \citep{peng2015}, from measurements of galaxy's hot gas halos \citep{wagner2018}, or indirectly through estimates of quenching times \citep[e.g.][]{taranu2014}.  The relevant timescale for quenching via starvation is the gas depletion time of a galaxy's present-day cold-gas reserves (since once this gas is consumed it will not be replenished), which is on the order of $\sim1-3\,\mathrm{Gyr}$ \citep{saintonge2017}.  Of course, it is likely that both ram pressure and starvation are acting in concert.  Observational studies have argued that starvation may drive an initial reduction in star formation with ram pressure ``finishing the job'' as galaxies approach the dense cluster centre \citep{vanderburg2018,roberts2019}.
\par
One technique for discriminating between quenching mechanisms is to study resolved properties of star formation within individual galaxies.  Ram pressure stripping preferentially removes gas from the outskirts of the galaxy where the gas is more loosely bound, which results in outside-in quenching.  Radio interferometers provide resolved maps of atomic and molecular hydrogen in galaxies and optical integral field unit (IFU) spectrographs provide resolved maps of common star formation tracers such as $\mathrm{H\alpha}$ emission.  With recent large optical IFU surveys such as MaNGA (Mapping Nearby Galaxies at Apache Point Observatory; \citealt{bundy2015}), CALIFA (Calar Alto Legacy Integral Field Area survey; \citealt{sanchez2012}), and SAMI (Sydney-AAO Multi-object Integral field spectrograph; \citealt{croom2012}), the number of galaxies with resolved \ha{} spectroscopy has increased rapidly.  Some of the galaxies observed by these surveys are located in dense, cluster environments and previous works have studied the resolved $\mathrm{H\alpha}$ properties in galaxy clusters.  Some studies have found evidence for outside-in quenching where the \ha{} profiles decrease rapidly with radius for cluster galaxies \citep[e.g.][]{schaefer2017,schaefer2019}.  Furthermore the GASP survey has identified many ``jellyfish galaxies'' in clusters which show tails of extended \ha{} emission \citep{poggianti2017}.  These jellyfish galaxies tend to be located in regions of the cluster where ram pressure forces are expected to be large \citep{jaffe2018}.  Additionally, many galaxies in nearby clusters have been observed to have extended HI tails \citep[e.g.][]{kenney2004, chung2007, chung2009, kenney2015}.  Ram pressure is expected to have a weaker effect on molecular gas ($\mathrm{H_2}$) which is more densely concentrated at the centre of galaxies.  The extent to which molecular hydrogen is stripped in galaxy clusters is still an open question \citep[e.g.][]{vollmer2012, jachym2019}.
\par
Candidates for galaxies undergoing stripping can also be identified using broad-band optical imaging.  Imaging in bluer filters in particular can efficiently highlight morphological features associated with ram pressure stripping \citep{mcpartland2015,poggianti2016}.  Follow-up observations of these disturbed cluster galaxies often show prominent ram pressure stripped tails of gas \citep{poggianti2017}. In this study we focus on the Coma cluster (Abell 1656), the nearest rich, high-mass galaxy cluster, to constrain the effects of ram pressure stripping on the population of satellite galaxies.  We take advantage of high-resolution, archival, multi-band imaging from the Canada-France-Hawaii Telescope (CFHT) to identify a sample of galaxies in the Coma cluster which appear to be experiencing stripping.  We then explore the observed properties of these ``stripping candidates'' and compare them to the rest of the Coma satellite population as well as isolated galaxies in the field.  The outline of this paper is as follows: in section~\ref{sec:data} we describe the Coma and field galaxy datasets, as well as describe the method we use to identify disturbed galaxies, potentially undergoing stripping; in section~\ref{sec:quant_morph} we compute quantitative morphological parameters for all Coma galaxies in order to compare to our visual classifications; in section~\ref{sec:orientation} we explore the orientation of observed asymmetries in stripping candidates with respect to the cluster centre; in section~\ref{sec:phase_space} we consider the position of stripping candidate galaxies in projected phase-space in order to constrain their infall histories; in section~\ref{sec:SFRmass} we compare star formation rates in stripping candidate galaxies compared to other Coma galaxies and field galaxies; and finally, in section~\ref{sec:disc_conc} we present the primary conclusions of this work and discuss these results.
\par
This paper assumes a flat $\mathrm{\Lambda}$ cold dark matter cosmology with $\Omega_M=0.3$, $\Omega_\Lambda=0.7$, and $H_0 = 70\,\mathrm{km\,s^{-1}\,Mpc^{-1}}$.  We assume a redshift for the Coma cluster of $z_\mathrm{coma} = 0.024$ and a luminosity distance to the Coma cluster of $105\,\mathrm{Mpc}$.

\section{Data}
\label{sec:data}

\subsection{Coma members}
\label{sec:coma_members}

We identify spectroscopic members of the Coma cluster using the twelfth data release of the Sloan Digital Sky Survey (SDSS-DR12, \citealt{alam2015}).  We consider Coma members to include any galaxies within $1 \times R_{200}$ and $3000\,\mathrm{km\,s^{-1}}$ of the cluster centroid, where $R_{200}$ is the virial radius of $2840\,\mathrm{kpc}$ \citep{kubo2007}.  This is a loose membership criteria which may select a small number of galaxies which are not strictly bound to the Coma cluster (in particular at large radius), however we opt for this approach to ensure that we do not miss galaxies that are just beginning their infall onto the cluster.  We use star formation rates (SFRs) and stellar masses ($M_\star$) from the medium-deep version of the GSWLC-2 SED fitting catalogue \citep{salim2016, salim2018}.  These SFRs are derived from UV+optical+mid-IR SED fitting done using the \textsc{cigale}\footnote{https://cigale.lam.fr/} code \citep{boquien2019}.  The IR SED is not explicitly fit (as it is not well constrained without far-IR information), instead the total IR luminosity (TIR) is estimated from templates using mid-IR fluxes and this TIR data point is used as a direct constraint in the SED fitting.
\par
We define star-forming galaxies to be all galaxies with $\mathrm{sSFR} > 10^{-11}\,\mathrm{yr^{-1}}$ ($\mathrm{sSFR} = \mathrm{SFR} / M_\star$) and passive galaxies to be all galaxies with $\mathrm{sSFR} \le 10^{-11}\,\mathrm{yr^{-1}}$.  This gives a sample of 296 star-forming Coma members and 388 passive Coma members.  We only consider satellite galaxies, and therefore exclude the two central, giant elliptical galaxies in Coma (NGC 4874 and NGC 4889) from our sample.  The median stellar mass for star-forming galaxies is $10^{9.1}\,\mathrm{M_\odot}$ and $10^{10.0}\,\mathrm{M_\odot}$ for passive galaxies.  The full range in stellar mass for all satellite galaxies is $M_\star = 10^{8.4} - 10^{11.3}\,\mathrm{M_\odot}$.  This sample is stellar mass complete down to roughly $\sim\!10^{8.8-9.0}\,\mathrm{M_\odot}$, which means that $\gtrsim\! 90$ per cent of the satellite galaxy sample is complete.  We note that if we restrict our sample to only galaxies with $M_\star > 10^9\,\mathrm{M_\odot}$, all of our conclusions are unchanged.  Finally, we calculate a rest-frame velocity dispersion for our Coma sample using the robust biweight estimator \citep{beers1990} and only considering the passive galaxy population - which is a better tracer of the cluster potential well \citep[e.g.][]{biviano1997,geller1999}.  This gives a velocity dispersion of $\sigma_\mathrm{los} = 930\,\mathrm{km\,s^{-1}}$, which is similar to previous estimates \citep[e.g.][]{colless1996}.

\subsection{Field galaxies}
\label{sec:field_sample}

For comparison, we also compile a sample of field galaxies from the SDSS.  We use the field sample described in \citet{roberts2017}, which is derived from $N=1$ ``groups'' in the \citet{yang2005, yang2007} SDSS DR7 group catalogue.  This isolated field sample is made up of all $N=1$ \citeauthor{yang2007} galaxies which are separated from their nearest ``bright'' neighbour by at least $1\,\mathrm{Mpc}$ and $1000\,\mathrm{km\,s^{-1}}$.  We restrict the sample to only include galaxies located within $3000\,\mathrm{km\,s^{-1}}$ of the Coma redshift, and we define a ``bright'' neighbour to be any galaxy which is brighter than the SDSS r-band absolute magnitude limit at $z=0.024$.  We also obtain SFRs and stellar masses for the field sample from the medium-deep GSWLC-2 catalog \citep{salim2016,salim2018}.  These cuts give an isolated field sample consisting of 3575 galaxies which are matched in redshift to the sample of Coma members. 

\subsection{Identifying stripping candidates with CFHT imaging}
\label{sec:cfht_imaging}

\begin{figure*}
    \centering
    \includegraphics[width=\textwidth]{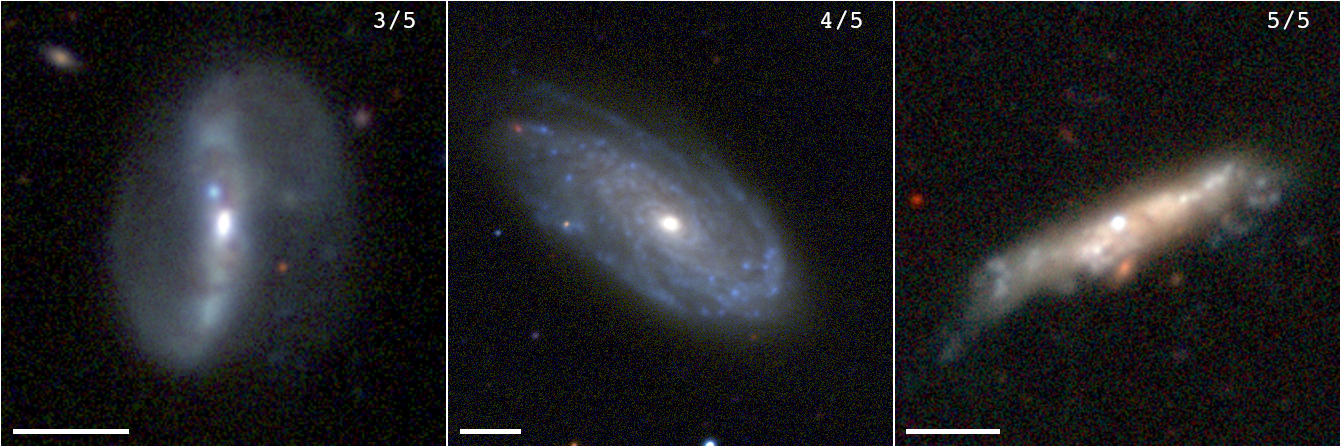}
    \caption{CFHT $ugi$ images for three identified stripping candidates.  We show galaxies which received 3/5, 4/5, and 5/5 votes from the classifiers.  The scale bar in each image corresponds to a physical size of 5 kpc.}
    \label{fig:mosaic_votes}
\end{figure*}

The primary goal of this work is to study the properties of galaxies potentially undergoing stripping in the Coma Cluster. To identify these ``stripping candidates'', we use high-quality, archival CFHT $ugi$ imaging (P.I. Hudson, Run ID 2008AC24) covering Coma out to the virial radius ($\sim 9\,\mathrm{deg}^2$), with exposure times of 300, 300, and 1360 seconds respectively.  We use image stacks produced by \textsc{megapipe} \citep{gwyn2008} which were downloaded from the CADC CFHT Science Archive\footnote{https://www.cadc-ccda.hia-iha.nrc-cnrc.gc.ca/en/cfht/}.  The average image quality of the stacks is $0.97\arcsec$, $0.85\arcsec$, and $0.73\arcsec$ for the $u$, $g$, and $i$ bands respectively.  \textsc{megapipe} also estimates the $5\sigma$ point source detection limit for which the average of the stacks is 26.2 mag, 25.5 mag, and 24.8 mag for the $u$, $g$, and $i$ bands respectively.  These magnitude limits are estimated simplistically by finding the faintest point source whose error is $0.198\,\mathrm{mag}$ or less.  The magnitude error is estimated as 
\begin{equation}
    \mathrm{mag_{err}} = 2.5 \, \log\,(1+N/S),
\end{equation}
\noindent
therefore for $S/N = 5$ this gives
\begin{equation}
    \mathrm{mag_{err}} = 2.5 \, \log\,(1.2) = 0.198.
\end{equation}
\noindent
In practice, this simple method gives magnitude limits which are accurate to $\sim\!0.3\,\mathrm{mag}$ \citep{gwyn2008}.
\par
To identify stripping candidates, we visually inspect CFHT three-colour $ugi$ images for all star-forming Coma member galaxies.  Colour cutout images are made with \textsc{stiff}\footnote{https://www.astromatic.net/software/stiff}\citep{bertin2012} covering a $40 \times 40\,\mathrm{kpc}$ box centred on each star-forming galaxy.  These images are then visually classified by five experts (including the authors of this work), all of whom are active researchers (see Acknowledgements) studying galaxy evolution with experience identifying galaxies undergoing stripping.  The classifiers were all given identical instructions to follow when classifying the images.  They were instructed to flag any galaxies which exhibited one, or more, of the following features:
\begin{enumerate}
    \itemsep0.5em
    \item \textit{The presence of asymmetric tails.}  Observed either in u-band (blue emission) or in dust (dark red extinction).
    \item \textit{Asymmetric star formation.}  u-band (blue) emission which is knotty and clearly asymmetric about the galaxy centre.
    \item \textit{The presence of bow shocks.}  Shock front features may be observed either in u-band (blue emission) or in dust (dark red extinction).
    \item \textit{Galaxy mergers.}  Either obvious mergers based on the presence of two interacting galaxies, or the presence of multiple, bright galaxy nuclei for the case of more evolved mergers.
\end{enumerate}
\noindent
Galaxies that were flagged as potential mergers by any of the classifiers were then discussed amongst the classifiers at a follow-up meeting.  A consensus was reached for each case regarding whether or not clear evidence of a galaxy-galaxy interaction was present.  All galaxies deemed to show clear evidence of a galaxy-galaxy interaction were then removed from the sample.  The final sample of stripping candidates is defined to include all galaxies flagged as hosting asymmetric tails and/or asymmetric star formation and/or shock features by a majority of the classifiers (ie. at least 3/5 classifiers).  This process results in 41 galaxies identified as stripping candidates out of a parent sample of 296 star-forming Coma members.  In the sample of stripping candidates, 17 per cent were identified by 3/5 classifiers, 27 per cent were identified by 4/5 classifiers, and 56 per cent were identified by all classifiers.  In Fig~\ref{fig:mosaic_votes} we show colour images for three example stripping candidates, one identified by 3/5 votes, one identified by 4/5 votes, and one identified by 5/5 votes. The $40 \times 40\,\mathrm{kpc}$ cutouts for all of the stripping candidates, along with a table of basic properties, are shown in Appendix~\ref{sec:image_appendix}.  Throughout this paper we will consistently refer to four different galaxy subsamples using the following nomenclature: 1. \textit{stripping candidates:} galaxies flagged by a majority of classifiers as potentially undergoing stripping, 2. \textit{star-forming Coma galaxies:} all star-forming Coma member galaxies ($\log \mathrm{sSFR} > -11$) not in the stripping candidate sample, 3. \textit{passive Coma galaxies:} all passive Coma member galaxies, 4. \textit{field galaxies:} galaxies in the isolated field sample.  The galaxies in the three Coma samples were visually inspected, and all merging/interacting galaxies were removed from the samples.
\par
We emphasize that some of our stripping candidates have been previously identified as galaxies undergoing stripping \citep[e.g.][]{yagi2007,smith2010b,yagi2010,yoshida2012,gavazzi2018,cramer2019}.  We identify 8/13 galaxies with UV tails from \citet{smith2010b} as stripping candidates in this work.  Of the remaining five, we flagged one as a potential post-merger (GMP 4555), two do not have SDSS spectroscopic redshifts (GMP 3016, GMP 4232), and two have SDSS DR12 redshifts inconsistent with the Coma Cluster, which prevented them from making it into our initial sample (GMP 2640, GMP 4060).  We identify 6/14 galaxies with $\mathrm{H\alpha}$ tails from \citet{yagi2010} as stripping candidates.  Of the remaining eight, five are in our Coma sample but were not identified as stripping candidates (GMP 2923, GMP 3071, GMP 3896, GMP 4017, GMP 4156) -- highlighting the fact that galaxies with $\mathrm{H\alpha}$ tails can have relatively undisturbed broad-band morphologies, two are lacking SDSS redshifts (GMP 3016, GMP 4232), and one has an SDSS DR12 redshift inconsistent with the Coma Cluster (GMP 4060).  Most previous studies have focused on the core of the Coma Cluster, therefore it is in that region where there is the most overlap with previous studies. Even with this overlap, the majority of galaxies in this paper are newly identified ram pressure stripping candidates.

\section{Quantitative measures of morphology}
\label{sec:quant_morph}

In section~\ref{sec:data} we described our procedure for visually identifying galaxies potentially undergoing stripping.  Visual classifications have their advantages when it comes to identifying specific morphological features associated with stripping, however these classifications are inherently qualitative in nature.  In this section we compute commonly used morphological measures, such as concentration-asymmetry and Gini-M20, in order to more quantitatively compare the morphologies of galaxies in this sample.  The high-quality CFHT images enable us to determine and compare quantitative morphologies between passive Coma members, star-forming Coma members, and stripping candidates.

\subsection{Creating segmentation maps}
\label{sec:seg_maps}

A challenge when computing morphological parameters is determining which pixels to include from the galaxy image.  This is particularly difficult in dense, low-redshift galaxy clusters where fields are often crowded by other cluster galaxies as well as background sources.  When computing morphologies we want to ensure, as much as possible, that we are only including pixels which are associated with the galaxy of interest. To do this we create segmentation maps beginning with the $40 \times 40\,\mathrm{kpc}$ cutouts centred on each galaxy from the CFHT imaging of the Coma cluster.  We then perform source detection on each cutout using the \texttt{detect\_threshold} and \texttt{detect\_sources} functions from the \textsc{Python} package \textsc{photutils}\footnote{https://photutils.readthedocs.io/en/stable/}.  We require that all source pixels be at least $2\sigma$ above the background level, where for each cutout a scalar background is estimated using sigma-clipped statistics.  We then generate a ``first-pass'' segmentation map for each cutout, requiring that all sources have at least five connected pixels above the background threshold.  For the majority of galaxies in our sample, this first-pass segmentation map provides an accurate description of the pixels associated with a given galaxy of interest.  However, a fraction of the identified stripping candidates show trails of stripped debris that can be physically separated from the main galaxy.  In this case, the image segmentation identifies this detached stripped material as sources separate from the main galaxy.  Since we are interested in describing the morphologies of galaxies undergoing stripping, it is important to include this material in the morphological calculations.  Therefore, for each stripping candidate, the first-pass segmentation map is visually inspected next to the $ugi$ thumbnail of the same galaxy.  Using the colour image as a reference, the first-pass segmentation maps are updated such that any distinct sources which appear to be stripped material are now given the same value as the main galaxy in the segmentation map.  Visual inspection showed that this amendment to the first-pass segmentation map was necessary for 13/41 stripping candidates.  These updated segmentation maps are then used to compute morphological parameters for the stripping candidates.  For the non-stripping galaxies, any mergers were identified and removed from the sample and for the remaining galaxies we find that the first pass segmentation maps were all sufficient.  We note that the qualitative results (Fig.~\ref{fig:quant_morph}) are unchanged when using first-pass segmentation maps for all of the stripping candidates instead of the segmentation maps which were updated manually.

\subsection{Morphology diagnostics}
\label{sec:morph_diagnostic}

\begin{figure*}
    \centering
    \includegraphics[width=\textwidth]{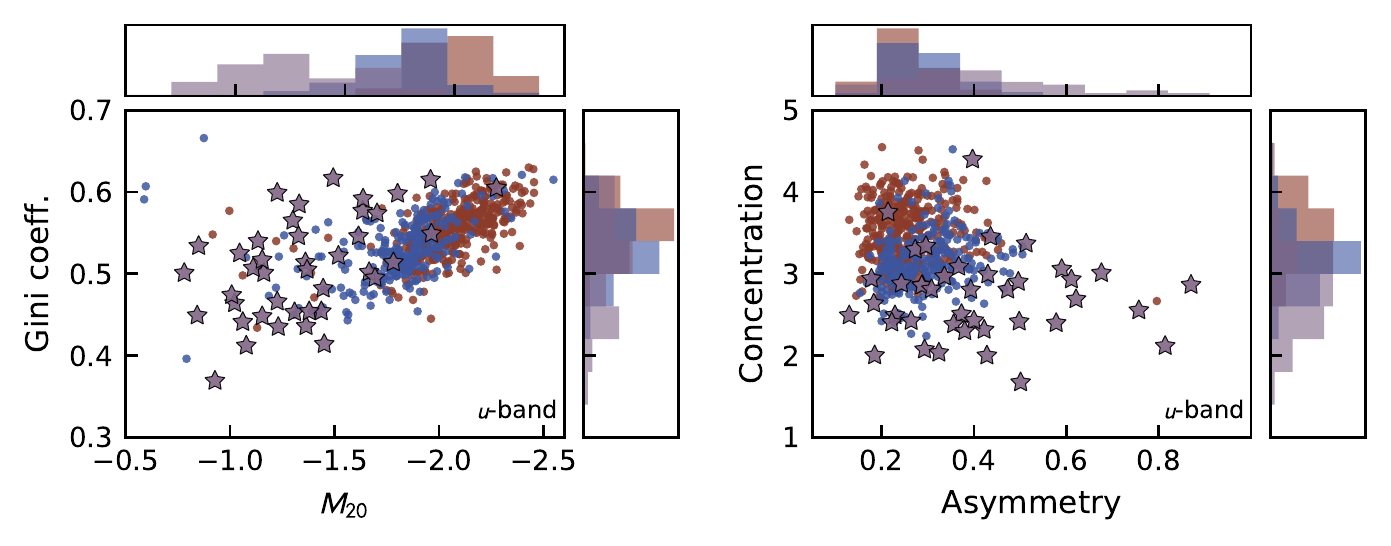}
    \caption{$u$-band Gini-$M_{20}$ diagram (left) and Concentration-Asymmetry diagram (right), for passive Coma galaxies (red), non-stripping star-forming galaxies (blue), and stripping candidates (purple).  Marginal distributions for each of the axes are shown with histograms.}
    \label{fig:quant_morph}
\end{figure*}

For all galaxies we compute a number of quantitative morphological parameters: the ``shape asymmetry'' and concentration, as well as the Gini coefficient and $M_{20}$.  Below we give a brief description of these morphological measures, however for a more detailed description please see the original papers \citep{abraham2003,lotz2004,conselice2003,conselice2014,pawlik2016}.  We use the \textsc{Python} package \textsc{statmorph} to compute all quantitative morphological parameters, for a full description of the implementation please see \citet{rodriguez-gomez2019}. \\[0.5em]
\noindent \textbf{\textit{Asymmetry:}}  We use the shape asymmetry ($A_S$) from \citet{pawlik2016} as our quantitative measure of asymmetry. The shape asymmetry is similar to the asymmetry index from the Concentration-Asymmetry-Clumpiness (CAS) system \citep{conselice2003,conselice2014}, however it is computed on a binary detection map (segmentation map) instead of the galaxy image.  The shape asymmetry is equivalent to a version of the CAS asymmetry which is not flux-weighted, and because of this, the shape asymmetry increases the sensitivity to low surface brightness features.  Because stripped galaxies often show low surface brightness tails or asymmetries we opt to use the shape asymmetry in this work, however we note that all of the qualitative morphological trends that we report are unchanged when using the standard flux-weighted CAS asymmetry as opposed to the shape asymmetry.  The shape asymmetry is computed as
\begin{equation}
    A_S = \mathrm{min}\,\left(\frac{\sum | X_0 - X_{180} |}{\sum |X_0|}\right)
\end{equation}
\noindent
where $X_0$ corresponds to the binary detection map and $X_{180}$ corresponds to the binary detection map rotated by $180^\circ$.  The rotation is done about the pixel which minimizes the standard (flux-weighted) asymmetry.  
\par
The binary detection map used to compute the shape asymmetry is generated as described in \citet{rodriguez-gomez2019}.  Briefly, a background level is first estimated over a circular annulus with inner and outer annuli of two and four times the Petrosian semimajor axis.  Then, a $1\sigma$ brightness threshold is defined and the galaxy image is smoothed with a $3\times 3$ boxcar (mean) filter.  The binary detection mask is then given by the contiguous group of pixels above the threshold that includes the brightest pixel in the galaxy image.  All following references to asymmetry will be referring to the shape asymmetry described above. \\[0.5em]
\noindent \textbf{\textit{Concentration:}}  The concentration parameter is defined as \citep{conselice2003,conselice2014}
\begin{equation}
    C = 5\,\log\,\left(\frac{r_{80}}{r_{20}}\right),
\end{equation}
\noindent
where $r_{80}$ is the radius containing 80 per cent of a galaxy's light and $r_{20}$ is the radius containing 20 per cent of a galaxy's light.  The total flux for each galaxy is taken to be the flux contained within $1.5\,r_\mathrm{petro}$ of the galaxy centroid. \\[0.5em]
\noindent \textbf{\textit{Gini coefficient:}}  The Gini coefficient originates from economics as a measure to quantify the distribution of wealth over a population.  However, it can also be applied to astronomical imaging data to quantify to homogeneity of flux distributed across galaxy pixels \citep{abraham2003,lotz2004}.  The Gini coefficient is computed as \citep{glasser1962}
\begin{equation}
    G = \frac{1}{\bar{X} n (n-1)} \sum_{i=1}^n (2i-n-1) X_i
\end{equation}
\noindent
where $i=1,2,3,...,n$ for a set of $n$ pixel flux values $X_i$.  A galaxy with all flux concentrated in one pixel corresponds to $G=1$, and a galaxy with a perfectly uniform flux distribution corresponds to $G=0$. \\[0.5em]
\noindent $\bm{M_{20}:}$  The $M_{20}$ statistic \citep{lotz2004} is a measure of the second-order moment of the galaxy image for the brightest 20 per cent of a galaxy's flux, normalized by the second-order moment for the entire galaxy image.  $M_{20}$ is particularly sensitive to bright features offset from the galaxy centre.  The ``total'' moment for the entire image is computed as
\begin{equation}
    \mu_\mathrm{tot} = \sum_{i=1}^N \mu_i = \sum_{i=1}^n f_i [(x_i - x_c)^2 + (y_i - y_c)^2]
\end{equation}
\noindent
where $x_i$, $y_i$ are the coordinates of the $i$th pixel, $f_i$ is the flux in the $i$th pixel, and $x_c$, $y_c$ are the central coordinates which minimize the total moment, $\mu_\mathrm{tot}$.  To compute $M_{20}$, galaxy pixels are rank-ordered by flux and $\mu_i$ is summed over the brightest pixels until the cumulative flux equals 20 per cent of the total flux.  This parameter is then normalized by the total moment, $\mu_\mathrm{tot}$
\begin{equation}
    M_{20} = \log\,\left(\frac{\sum_i \mu_i}{\mu_\mathrm{tot}}\right),\; \mathrm{while}\;\sum_i f_i < 0.2\,f_\mathrm{tot},
\end{equation}
\noindent
where $f_\mathrm{tot}$ is the total flux of the pixels identified by the segmentation map.
\par
In Fig.~\ref{fig:quant_morph} we plot Coma member galaxies (passive: red, star-forming: blue, visual stripping candidates from section~\ref{sec:cfht_imaging}: purple) on the Gini-M20 (left) and concentration-asymmetry (right) diagrams using the calculated morphological parameters.  In both cases there is a clear separation between the star-forming/passive Coma galaxies and the stripping candidates.  For each panel in Fig.~\ref{fig:quant_morph} we also include 1D histograms corresponding to each axis.  Stripping candidates have lower Gini coefficients, less-negative values of $M_{20}$, lower concentrations, and larger asymmetries compared to other Coma galaxies.  The separation between star-forming/passive Coma galaxies and stripping candidates is largest when considering $M_{20}$ or asymmetry.  Stripping candidates were selected based on visible signs of asymmetric star formation, asymmetric tails, and shock-front features, therefore it is unsurprising that they show preferentially large measured asymmetries; however it is a reassuring confirmation of the visual classifications.  The values of $M_{20}$ for stripping candidates are consistent with a relatively diffuse population of galaxies, which is in turn confirmed by the low concentrations.  Furthermore, the fact that $M_{20}$ values for stripping candidates are closer to zero may be driven in part by bright features offset from the galaxy centre in stripping candidates, such as shock fronts or stripped tails/knots of star formation.  \citet{mcpartland2015} have performed a similar morphological analysis of galaxies undergoing stripping using \textit{HST} observations at intermediate redshift ($z=0.3-0.7$), finding that ram pressure stripping candidates at intermediate redshift occupy distinct regions in the $G-M_{20}$ and $C-A$ planes.  Here we find that the morphological results from \citet{mcpartland2015} at $z=0.3-0.7$ are consistent with these results from ground-based imaging at low-redshift.

\section{Orientation of stripping features}
\label{sec:orientation}

\begin{figure}
    \centering
    \includegraphics[width=0.9\columnwidth]{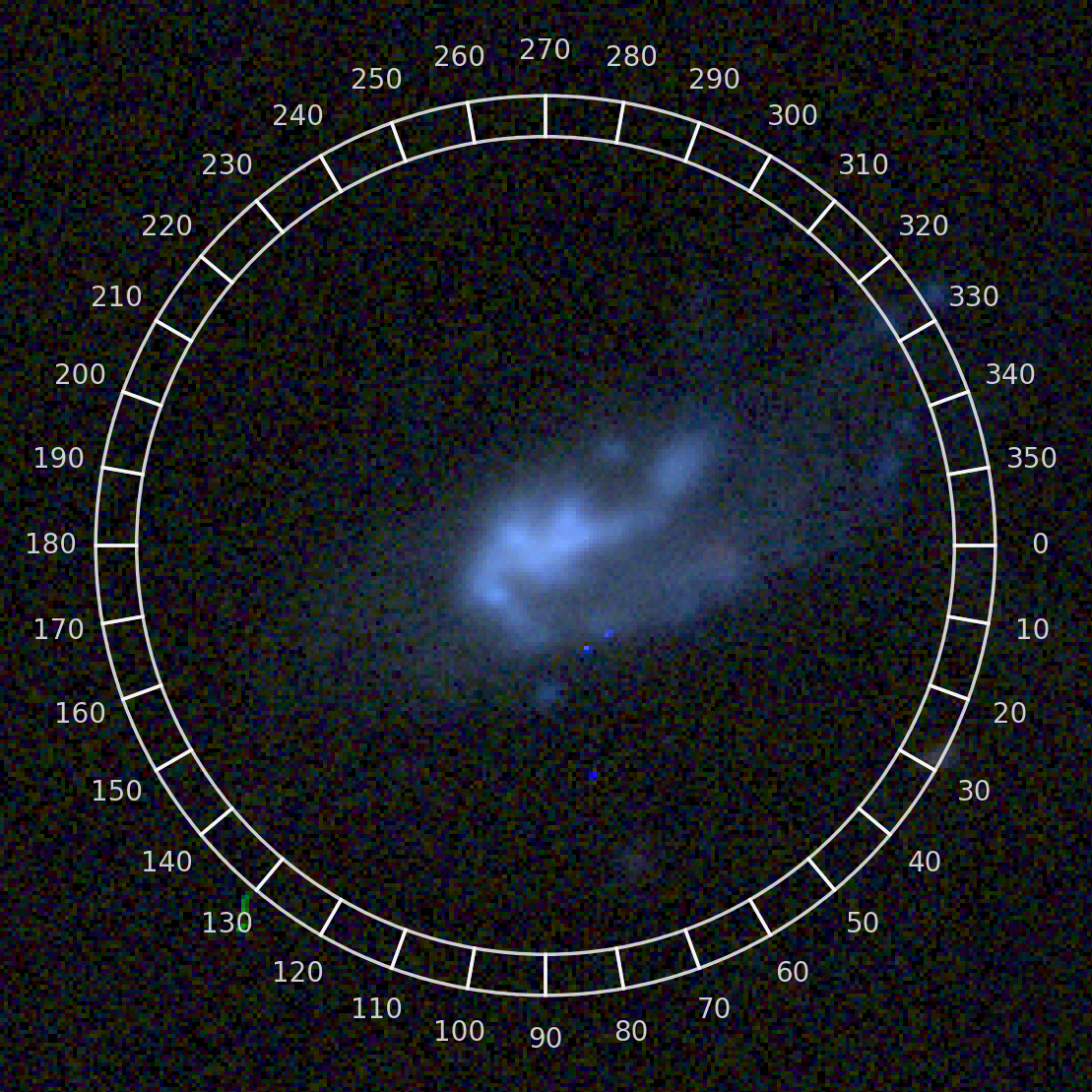}
    \caption{Example of $ugi$ thumbnail with angular guide overlaid in order to determine orientation of observed stripping features.}
    \label{fig:rgb_ang}
\end{figure}

\begin{figure*}
    \centering
    \includegraphics[width=\textwidth]{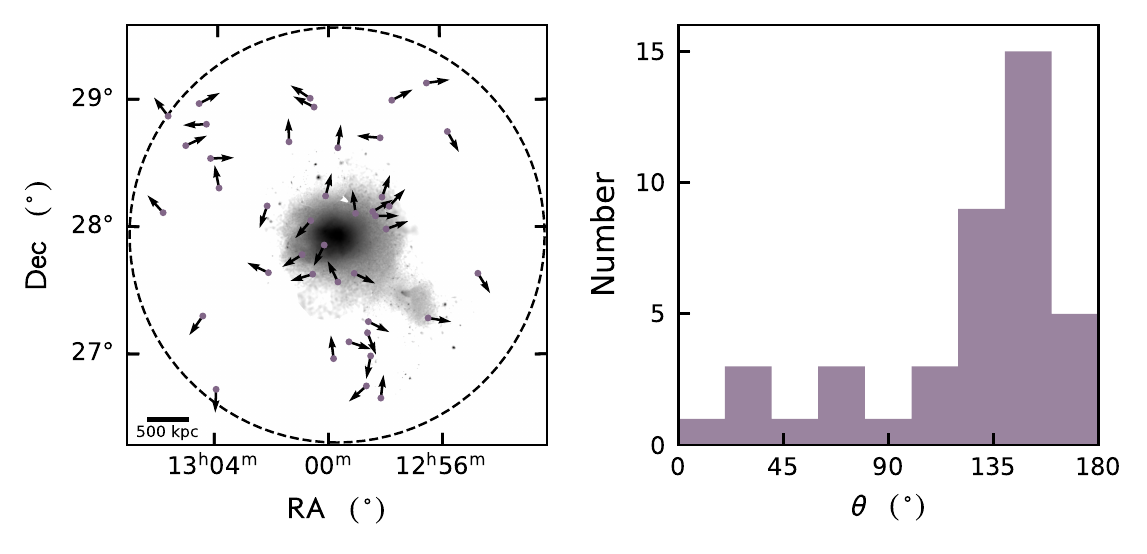}
    \caption{\textit{Left:}  Vector plot showing the orientation at the identified ``stripping features'' within the Coma cluster.  Each point corresponds to a stripping candidate and the arrow points in the direction of the stripping feature, as estimated by eye (see text).  The dashed line marks the virial radius, and in the background grayscale we show an 0.5-7 keV XMM-Newton X-ray image.  Note that the XMM-Newton observations only cover the inner $\sim\! 1000-1500\,\mathrm{kpc}$ of Coma.  \textit{Right:}  Histogram of the angle between the stripping feature direction and the direction toward the cluster centre.  A majority of galaxies show angles between $\sim\!120^\circ$ and $180^\circ$, pointing away from the cluster centre.}
    \label{fig:asym_angle}
\end{figure*}

A key signature of ram pressure stripping are tails of gas and/or stars trailing behind galaxies opposite to the direction of motion.  These ram pressure tails have been observed extensively in cluster galaxies across the electromagnetic spectrum \citep[e.g.][]{kenney2004,smith2010b,jachym2014,poggianti2017,cramer2019b}.  Simulations and observations show that galaxies infall onto clusters on largely radial orbits \citep[e.g.][]{wetzel2011,biviano2013,lotz2019}, meaning that tails pointing towards or away from the cluster centre are expected for galaxies undergoing strong stripping.  Tails pointing away from the cluster centre indicate galaxies infalling towards the cluster centre; conversely tails pointing towards the cluster centre are suggestive of galaxies ``backsplashing'' away from the cluster centre after a pericentric passage.
\par
For candidates identified as undergoing stripping, we visually estimate the direction of the stripping features relative to the galaxy centre.  In Fig.~\ref{fig:rgb_ang} we show an example of a thumbnail used to determine the orientation of the stripping features.  These are the same thumbnails used for visual classifications, however now a guide is overlaid marking angles between $0^\circ$ and $350^\circ$.  The orientation of the stripping features is determined by selecting the angle (in multiples of ten degrees) which is most consistent with the observed features.  These angles are determined for each galaxy by the five expert classifiers, and the final ``asymmetry angle'' is taken to be the average of the five estimates for each galaxy.  We note that for galaxies where the identified features are relatively subtle, identifying these orientations can be difficult and subjective.  By averaging over five classifiers we help to mitigate these difficulties, however some level of subjectivity will persist.  For more than half of the cases, there was very good agreement (little-to-no scatter) in the angle estimates.  If we only consider galaxies where the selected angles were relatively consistent between the classifiers (ie. $\sigma / \sqrt{N}$ errors $<\!15^\circ$), the qualitative results presented below are unchanged compared to using the entire sample.  For the example shown in Fig.~\ref{fig:rgb_ang}, an average angle of $336^\circ$ was determined through this process.  For each thumbnail in appendix~\ref{sec:image_appendix}, we overlay an arrow showing the orientation of the stripping features for each galaxy.
\par
In Fig.~\ref{fig:asym_angle} (left) we show the spatial distribution of stripping candidates within the Coma cluster.  The plotted vectors point in the direction of the observed ram pressure features.  In the background we plot a 0.5-7 keV X-ray image of the Coma Cluster from \textit{XMM-Newton}, where the main Coma X-ray peak, as well as the secondary peak corresponding to the infalling NGC 4839 group, are both visible.  The X-ray data were reduced, imaged, and mosaiced following the standard procedure outlined in the \textit{XMM-Newton} ESAS cookbook\footnote{https://heasarc.gsfc.nasa.gov/docs/xmm/esas/cookbook/xmm-esas.html} (see \citealt{roberts2018} for a detailed description of the reduction process).  It is clear by eye that the majority of the vectors are pointing away from the cluster centre, which is consistent with the majority of these galaxies being stripped during infall toward the cluster centre. With these estimated directions, we compute the angle between the direction of the stripping feature and the centre of the cluster.  An angle of $0^\circ$ corresponds to a tail pointing directly toward the cluster centre, and an angle of $180^\circ$ corresponds to a tail pointing directly away from the cluster centre.  In Fig.~\ref{fig:asym_angle} (right) we show a histogram of the angle between the stripping feature and the cluster centre.  The majority of stripping galaxies have angles between $120^\circ$ and $180^\circ$.  Naively, if these morphological features were not associated with ram pressure then we would expect to see a uniform distribution of angles instead of the clearly preferred angles seen in Fig.~\ref{fig:asym_angle}.  For satellites on perfectly radial orbits, ram pressure features should be oriented at angles to the cluster centre of $0^\circ$ or $180^\circ$.  While simulations predict that satellites infall on largely radial orbits \citep[e.g.][]{wetzel2011}, perfectly radial orbits are an overly simplistic assumption and most galaxy orbits have a non-negligible tangential component \citep[e.g.][]{biviano2013}.  Both variations in orbits and projection effects make it difficult to interpret the observed tail directions.  Some ram pressure studies have found strongly peaked distributions of angles \citep[e.g.][]{chung2007,smith2010b}, while others have found more random distributions \citep[e.g.][]{mcpartland2015,poggianti2016}.  In a simplistic picture, the distribution of angles observed in this work is consistent with expectations from ram pressure stripping, and the fact that we do not observe a peak at precisely $180^\circ$ is consistent with orbits which are slightly non-radial on average.  Orientations measured in projection will always be inherently uncertain, which must be remembered when interpreting the results in Fig.~\ref{fig:asym_angle}

\section{Phase space analysis}
\label{sec:phase_space}

\begin{figure}
    \centering
    \includegraphics[width=\columnwidth]{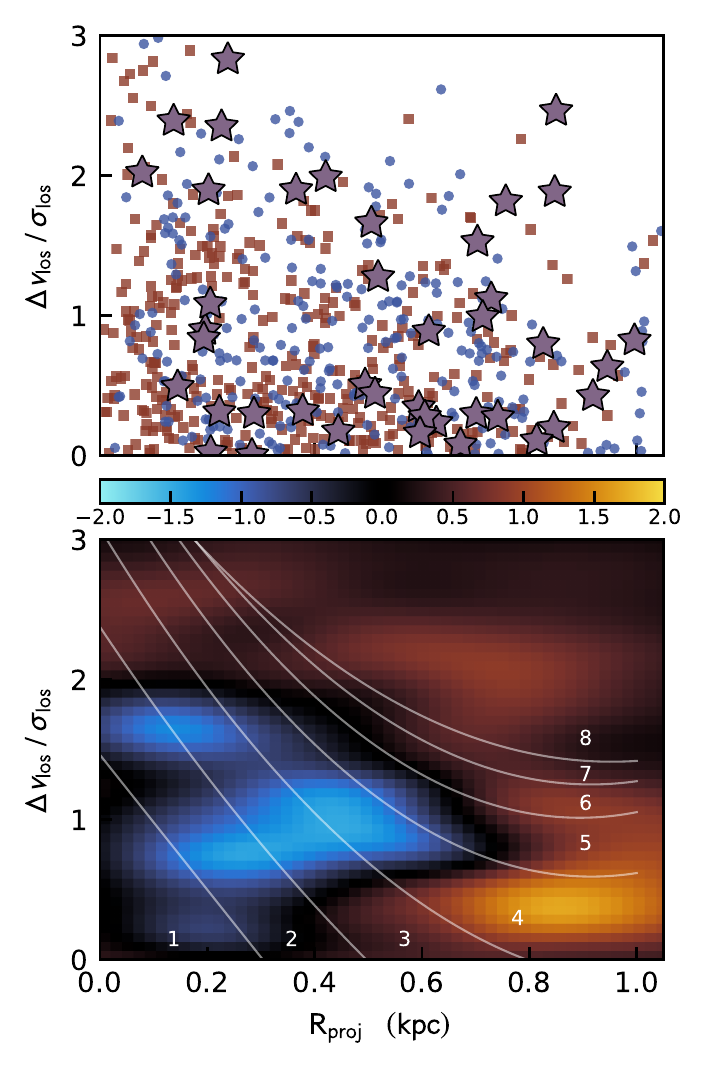}
    \caption{\textit{Top:}  Projected phase space diagram for member galaxies of the Coma cluster.  Passive galaxies are shown as red squares, star-forming galaxies are shown as blue circles, and stripping candidates are shown as purple stars.  \textit{Bottom:}  Difference between the phase space KDE distributions for stripping candidates and normal star-forming Coma galaxies.  The colorbar corresponds to the fractional change relative to the mean kernel density.  At all radii, there are an excess of stripping candidates (red colour) at large velocity offsets, relative to the bulk star-forming population (blue colour).}
    \label{fig:phase_space}
\end{figure}

An observational tool to study the accretion history of galaxy clusters is the projected phase space (PPS) diagram, which plots 1D velocity offset versus projected cluster-centric distance for member galaxies. Galaxy distributions in PPS tend to trace caustics corresponding to escape velocities from the the host cluster halo.  At small cluster-centric distance the velocity range is large due to the high escape velocity, whereas at large distance, where the escape velocity is lower, the velocity range narrows.  Furthermore, galaxies first infalling onto a galaxy cluster follow distinct orbits in phase space.  They are accelerated to large velocity offsets on their first infall toward the cluster centre, and then over the course of multiple orbits the velocities approach the cluster centroid (see e.g. Fig. 1 in \citealt{rhee2017}).  This means that galaxies on first infall tend to have large velocity offsets, often near the escape velocity caustic.  These infalling tracks in phase space are clear in simulations when using full 3D positions and velocities.  In projection, these infalling tracks are far less clear, however infalling galaxies still tend to be found at the velocity outskirts of PPS \citep[e.g.][]{mahajan2011,oman2013}.
\par
In Section~\ref{sec:orientation} we showed that the orientation of stripping features is consistent with the majority of stripping candidates being on first infall toward the cluster centre.  We further test this by considering the position of stripping candidates in PPS.  In Fig.~\ref{fig:phase_space} (top) we plot the PPS diagram for star-forming (blue) and passive (red) Coma members, as well as stripping candidates (purple stars).  There is a collection of stripping candidates extending from $\Delta v_\mathrm{los} \sim 0.5\,\sigma_\mathrm{los}$ at the virial radius to $\Delta v_\mathrm{los} \sim 3\,\mathrm{\sigma_{los}}$ near the cluster centre.  This is consistent with the PPS positions expected for infalling galaxies.  To further compare the distribution of stripping candidates in PPS to that for star-forming Coma galaxies, we measure a 2D gaussian kernel density estimate (KDE) for the PPS distribution for star-forming galaxies and stripping candidates.  In Fig.~\ref{fig:phase_space} (bottom) we show the difference between the KDE distributions for the stripping candidates and the non-stripping star-forming galaxies.  Blue regions in this map correspond to an excess of non-stripping galaxies and red/orange regions correspond to an excess of stripping candidates.  It is clear that relative to the non-stripping population, the stripping candidates are preferentially found along the infalling track in PPS.  This is further evidence that many of these galaxies are being stripped on their first infall toward the cluster centre.  At almost all cluster-centric radii, stripping candidates are preferentially found at large velocity offsets.  Stripping candidates should therefore be experiencing a stronger ram pressure force (scales with $\sim\!\rho v^2$), which is further evidence that it is ram pressure stripping driving the observed asymmetries in these galaxies.  These results are in qualitative agreement with \citet{jaffe2018} who find that GASP jellyfish galaxies are preferentially found at large peculiar velocities within their host clusters.

\begin{figure}
    \centering
    \includegraphics[width=\columnwidth]{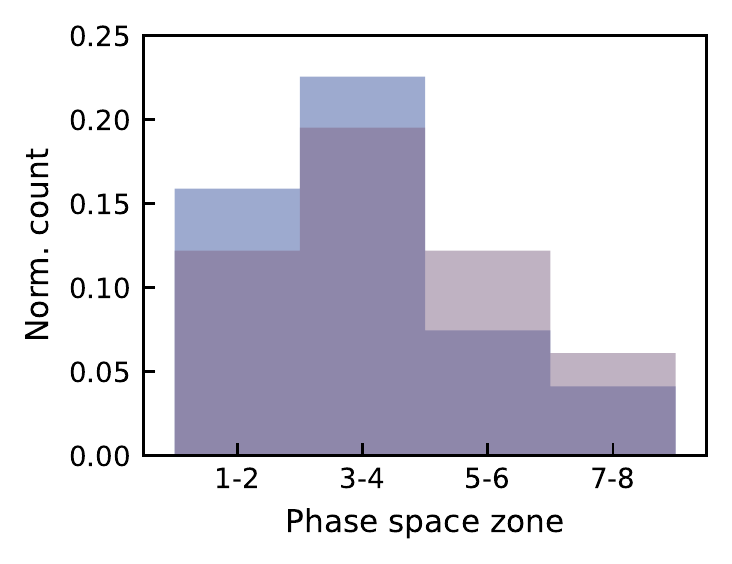}
    \caption{Normalized histograms showing the distribution of phase space zones from \citet{pasquali2019} for star-forming Coma members (blue) and stripping candidates (purple).  These phase-space zones trace time-since infall, with the average time-since infall increasing monotonically from zone 8 (1.42 Gyr) to zone 1 (5.42 Gyr).}
    \label{fig:PSzone}
\end{figure}

Using the Yonsei suite of galaxy cluster zoom simulations (in projection), \citet{pasquali2019} derived regions in PPS of roughly constant time-since-infall.  These zones allow us to quantitatively constrain, at least on average, the infall times of Coma galaxies. In Fig.~\ref{fig:phase_space} (bottom) we show these PPS regions numbered from 1-8.  The average time-since-infall for simulated galaxies in these zones from \citet{pasquali2019} increase monotonically from 1.42 Gyr in zone 8 to 5.42 Gyr in zone 1.  In Fig.~\ref{fig:PSzone} we plot normalized histograms distribution of PPS zones of star-forming Coma galaxies and stripping candidates, showing an excess of stripping candidates in the higher PPS zones, which correspond to shorter times-since-infall.  This quantitatively demonstrates that many stripping candidates are likely recent infallers.

\section{Star formation in stripping galaxies}
\label{sec:SFRmass}

\begin{figure*}
    \centering
    \includegraphics[width=\textwidth]{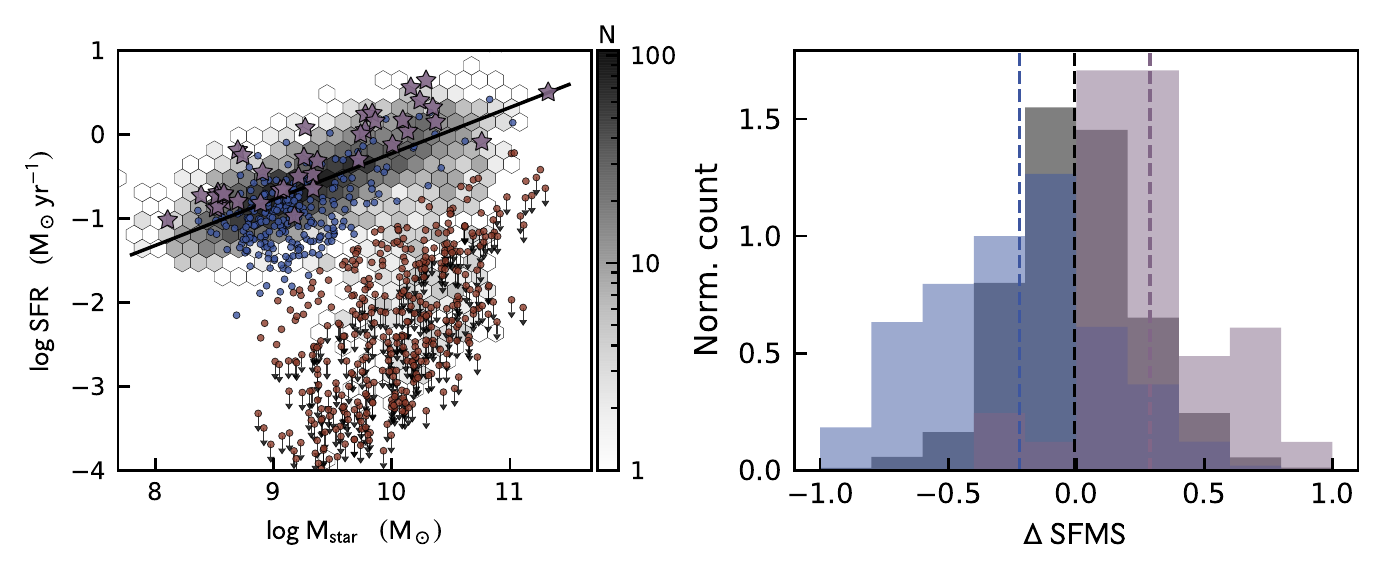}
    \caption{\textit{Left:}  Star formation rate versus stellar mass.  Background greyscale shows distribution for galaxies from the isolated field sample and the trend line shows the best-fit to star-forming ($\mathrm{sSFR} > 10^{-11}\,\mathrm{yr^{-1}}$) field galaxies.  Blue points correspond to normal star forming Coma galaxies, red points show passive Coma galaxies, and purple stars denote Coma stripping candidates.  \textit{Right:}  Offset from the field star-forming main sequence for non-stripping star-forming galaxies (blue), field galaxies (black), and stripping candidates (purple).  Dashed lines show median offsets from the star-forming main sequence for each population.}
    \label{fig:SFRmass_field}
\end{figure*}

The effect of ram pressure stripping on galaxy star formation has been a topic of focus of many previous studies \citep[e.g.][]{quilis2000,steinhauser2016,poggianti2017,roberts2019}.  Ram pressure stripping is a mechanism which quenches star formation in cluster galaxies through rapid gas removal, however recent simulations and observations have shown that star formation may be briefly enhanced during stripping \citep{steinhauser2012,bekki2014,ebeling2014,poggianti2016,troncoso-iribarren2016,vulcani2018}, prior to gas removal.  The star formation enhancement is likely driven by gas compression due to shocks from ram pressure which catalyzes strong star formation.  The sample of galaxies in this work allow us to further test this prediction.
\par
In Fig.~\ref{fig:SFRmass_field} (left) we plot star formation rate versus stellar mass for isolated field galaxies (grey 2D histogram), all SDSS Coma galaxies (circles), and stripping galaxies (purple stars).  Blue circles correspond to star-forming galaxies and red circles correspond to passive galaxies.  We distinguish between star-forming and passive galaxies using a single cut in specific star formation rate at $\mathrm{sSFR} = 10^{-11}\,\mathrm{yr^{-1}}$.  In Fig.~\ref{fig:SFRmass_field} we mark the SFRs for all galaxies with $\mathrm{sSFR} < 10^{-11.7}\,\mathrm{yr^{-1}}$ as upper limits as suggested by \citet{salim2016}.  Finally, for reference we determine a star-forming main sequence (SFMS) by fitting a single powerlaw to the SFR-mass relationship for star-forming field galaxies.  We incorporate uncertainties in both stellar mass and SFR to fit the SFMS using \textsc{linmix}\footnote{https://linmix.readthedocs.io/en/latest/index.html} \citep{kelly2007}.  We find a main sequence relationship of
\begin{equation}
    \log\,\mathrm{SFR} = 0.55 \times \log M_\mathrm{star} - 5.7
\end{equation}
\noindent
which is shown in Fig.~\ref{fig:SFRmass_field} (left) with the solid black line.  As expected, the red-sequence is significantly more populated in the Coma cluster relative to the field, furthermore, star-forming Coma galaxies fall slightly below the SFMS suggesting that the cluster environment has an effect even on star-forming Coma galaxies.  Consensus is lacking regarding whether the slope and normalization of the SFMS depend on environment.  \citet{peng2010} have found that the SFMS is independent of environment (traced by local galaxy density), whereas other studies have found small offsets, $\sim\! 0.1-0.2\,\mathrm{dex}$, between the SFMS in the field versus groups or clusters \citep{vulcani2010,lin2014,erfanianfar2016,paccagnella2016,grootes2017,wangL2018}.  Such an offset may also only be present at low-redshift and not in the early Universe \citep{erfanianfar2016}.  The results of this work are consistent with previous studies which find an offset; we measure a small offset of $\sim\! 0.3\,\mathrm{dex}$ between the isolated field and Coma cluster SFMS.
\par
At all stellar masses the stripping candidates are found to have significantly higher SFRs relative to the bulk star-forming population in Coma (blue circles).  This is consistent with predictions that ram pressure can induce temporary enhancements in star formation prior to quenching.  When compared to field galaxies the stripping candidates still show enhanced SFRs, with 90 per cent of stripping galaxies falling above the field SFMS.  As a test, we randomly draw 41 galaxies (the size of the stripping candidate sample) from the star-forming field population and compute their offsets from the SFMS.  We then repeat this Monte-Carlo trial $100\,000$ times to determine what fraction of galaxies are scattered above the SFMS due to random chance alone.  Fractions above the SFMS of $>$70 per cent from our random samples only occur in 0.05 per cent of trials.  This is strong evidence that the SFRs of stripping candidates are systematically enhanced, both relative to other Coma galaxies and relative to isolated field galaxies.
\par
In Fig.~\ref{fig:SFRmass_field} (right) we show histograms of the offset from the field SFMS for star-forming Coma galaxies (blue), field galaxies (black), and stripping candidates (purple).  As previously stated, star-forming Coma galaxies fall preferentially below the field SFMS and stripping candidates fall preferentially above the SFMS.  Based on a $k$-sample Anderson-Darling test \citep{scholz1987}, a cumulative distribution test which tests the null hypothesis that k-samples are drawn from the same underlying distribution, these three distributions are distinct at greater than $99.99$ per cent confidence.  Quantitatively, the median offset from the SFMS is enhanced for stripping candidates by 0.5 dex relative to Coma star-forming galaxies and by 0.3 dex relative to field galaxies.  This offset from the SFMS of 0.3 dex for stripping candidates is similar to the star formation enhancement of 0.2 dex relative to the SFMS reported by \citet{vulcani2018} for GASP Jellyfish galaxies.

\section{Discussion and Conclusions}
\label{sec:disc_conc}

\subsection{Are ``stripping-candidates'' undergoing ram pressure stripping?}
\label{sec:strip_RPS?}

The sample of stripping candidates presented in this work were identified based on disturbed, asymmetric morphological features.  These visual signatures can be generated by stripping processes or possibly other interactions, such as harassment, tidal effects, or mergers.  As summarized below, we find that the population of stripping candidates is consistent with galaxies undergoing ram pressure stripping, however we note that some individual galaxies may be exceptions with different origins for their disturbed morphologies.
\par
In Fig.~\ref{sec:orientation} (right) we show the orientation of stripping features with respect to the cluster centre.  The observed distribution of angles, with one prominent peak pointing roughly away from the cluster centre, is consistent with simple expectations from ram pressure stripping.  We emphasize that the visual classifiers were given no information regarding positions within the Coma cluster when galaxies were visually classified, therefore this trend is not being driven by any selection biases in the classification process.  As in Section~\ref{sec:orientation}, we emphasize that the interpretation of observed asymmetry orientations is complicated by variations in galaxy orbits as well as projection effects.  The observed asymmetries in this work are broadly consistent with a simple picture of ram pressure stripping of galaxies on first infall.  Additionally, Fig.~\ref{fig:phase_space} shows that there are an excess of stripping candidates located at large velocity offsets relative to other Coma satellites.  Given that ram pressure scales with $\rho v^2$, this suggests that stripping candidates are currently experiencing a relatively strong ram pressure force.  Other cluster processes, such as tidal effects, which could give rise to asymmetric morphologies, should be occurring in regions where galaxy number densities are large.  We measure nearest-neighbour number densities for all of the galaxies in this work, and find that stripping candidates actually have marginally \textit{lower} nearest-neighbour number densities (at fixed cluster-centric radius) compared to star-forming Coma galaxies (plot not shown, significant at $2-3\,\sigma$).  This suggests that tidal interactions due to densely populated local environments is likely not the driving factor behind the observed morphologies of stripping candidates.
\par
Follow-up observations of these stripping candidates are essential to confirm (or rule out) ram pressure as the driver of their disturbed morphologies.  These stripping candidates have been identified largely on the basis of their stellar morphologies, which corresponds to a fairly tightly bound galaxy component.  Resolved observations of components more susceptible to stripping, such as atomic hydrogen or ionized gas traced by H$\alpha$, would provide even more information on the impacts of ram pressure on these galaxies \citep[e.g.][]{kenney2004,poggianti2017}.

\subsection{Star formation activity throughout ram pressure stripping}
\label{sec:disc_SF_RPS}

Assuming that the stripping candidates identified here represent galaxies undergoing ram pressure stripping, this sample provides important constraints on galaxy star formation throughout the stripping process.  In Fig.~\ref{fig:SFRmass_field} we show that the SFRs of stripping candidates are clearly enhanced relative to both star-forming Coma galaxies, as well as star-forming galaxies in the field.  This is consistent with previous observational work which has found similar enhancements of SFRs in galaxies experiencing ram pressure stripping \citep{ebeling2014,poggianti2016,vulcani2018}, as well as theoretical predictions from hydrodynamic simulations \citep{steinhauser2012,bekki2014,troncoso-iribarren2016}.  We note that the SFRs used in this work are derived from UV+optical+TIR SED fitting \citep{salim2016,salim2018}, however we also explore dust-corrected $\mathrm{H\alpha}$ fluxes from the SDSS spectra \citep{thomas2013}.  We find that the median $\mathrm{H\alpha}$ flux for stripping candidates is enhanced over the median $\mathrm{H\alpha}$ flux for star-forming field galaxies by a factor of $\sim\!2$.  When considering $\mathrm{H\alpha}$ fluxes we see an even stronger enhancement for stripping candidates over star-forming Coma galaxies.  This demonstrates that the observed star formation enhancement for stripping candidates is not only limited to the galaxy disc as a whole, but is also present in galaxy centres traced by the $3\arcsec$ SDSS fibre ($\sim\!1.5\,\mathrm{kpc}$ at the redshift of Coma).  Furthermore, star formation in stripped tails can make-up a non-negligible portion of the galaxy star formation budget \citep{poggianti2019}, and such extended star formation is likely not captured by the indicators used in this work.  Therefore, we may actually still be underestimating the enhancement of star formation in stripping candidates.
\par
After this period of enhanced star formation, the details of the ``quenching phase'' associated with ram pressure stripping depends strongly on the efficiency of ram pressure stripping.  For example, in the case of extremely efficient stripping, all (or most) of a galaxies atomic and molecular gas reserves may be directly stripped leading to rapid quenching.  However, the molecular gas component may be difficult to strip directly, as it is centrally concentrated and more strongly bound to the the host galaxy.  In this case, ram pressure stripping may be able to remove large amounts of atomic hydrogen but leave large molecular gas reserves unstripped.  In this case the quenching timescale would then be set by the depletion time of the remaining gas \citep[see e.g.][]{roberts2019}.  This scenario predicts the existence of a ``post-stripping" phase where galaxies show residual star formation along with a truncated gas disc due to stripping.  Such post-stripping galaxies have been observed in galaxy clusters \citep[e.g.][]{yoon2017,jaffe2018}.  The non-stripping population of star-forming galaxies that we identify in Coma are potentially a mixture of these post-stripping galaxies as well as normal star-forming galaxies that have not been strongly affected by ram pressure.  If some of these galaxies have begun to quench, that could explain the population of star-forming Coma galaxies located below the SFMS (see Fig.~\ref{fig:SFRmass_field}).  With this dataset we cannot measure the fraction of galaxies which may be post-stripping, as that requires resolved maps of star formation or gas in these systems.  We can, however, derive rough constraints on the timescales over which galaxies show morphological features of stripping, based on the fact that the vast majority of stripping candidates show morphological features that are pointing away from the cluster centre.  There seem to be very few stripping candidates that are on their way out of the cluster centre after a pericentric passage, which suggests that the period over which Coma galaxies show morphological signatures of stripping cannot last much longer than a crossing time.  We estimate the crossing time for Coma as $t_\mathrm{cross} \sim R_\mathrm{vir} / \sigma_\mathrm{los}$ which, for $R_\mathrm{vir}=2840\,\mathrm{kpc}$ and $\sigma_\mathrm{los}=930\,\mathrm{km\,s^{-1}}$, gives a crossing time of $\sim\!3\,\mathrm{Gyr}$.  Therefore, we can infer an upper limit for the period of strong ram pressure stripping of $\lesssim 3\,\mathrm{Gyr}$.  Further constraints on this timescale requires knowledge of when ram pressure stripping started for each galaxy, which varies depending on galaxy mass, orbits, gas distributions, etc.

\subsection{Identifying stripping galaxies with rest-frame optical imaging}
\label{sec:disc_imaging}

In this work we visually identify galaxies potentially undergoing stripping using three-colour rest-frame optical imaging.  We confirm the validity of these classifications by measuring quantitative morphological parameters for stripping candidates, and show that they occupy unique regions of commonly used morphological planes (see also \citealt{mcpartland2015}).  Based on Fig.~\ref{fig:quant_morph}, simple cuts in asymmetry or $M_{20}$ select the majority of stripping candidates, with minimal contamination from ``normal'' cluster galaxies, in an automated fashion.  In the era of wide-field photometric surveys, this is a potentially useful way to identify large numbers of candidate stripping galaxies.  While such an automated identification also flags mergers and other highly disturbed galaxies, in the cluster environment it is likely that stripping galaxies outnumber mergers.  At the very least, such a selection could narrow a prohibitively large sample for follow-up visual classifications.  In Fig.~\ref{fig:quant_morph} we show morphological parameters computed using $u$-band images, as we find the $u$-band provides the clearest separation between stripping candidates and other cluster galaxies, however we note that we still find a clear separation in the Gini-$M_{20}$ and concentration-asymmetry using $g$- or $i$-band images.

\subsection{Conclusions}

In this paper we present a sample of 41 galaxies visually identified as candidates for galaxies undergoing ram pressure stripping in the Coma cluster.  While some of these stripping-candidates have been previously identified, the majority of our sample are newly identified ram pressure candidate galaxies.  This sizable sample enables a detailed study of the properties of Coma galaxies experiencing stripping.  The main conclusions of this work are:

\begin{enumerate}
    \itemsep0.5em
    
    \item Stripping candidates are clear outliers, relative to normal cluster galaxies, in common morphology parameter spaces such as Gini-$M_{20}$ and concentration-asymmetry.
    
    \item Morphological stripping features (e.g. tails, asymmetric star formation, shock fronts) are preferentially oriented radially away from the cluster centre, with a minority of stripping candidates displaying features directed toward the cluster centre.  Virtually no stripping candidates show morphological features perpendicular to the cluster centre.
    
    \item The population of stripping candidates is consistent with most galaxies being on first infall toward the cluster centre.
    
    \item Star formation rates of stripping candidates are clearly enhanced, both relative to other star-forming Coma galaxies, and relative to isolated star-forming field galaxies.  This is consistent with ram pressure driving an enhancement in galaxy star formation.
\end{enumerate}

\noindent
Follow-up observations are essential to confirm the origin of these disturbed galaxies.  If confirmed as galaxies in the process of being stripped, that would suggest that ram pressure stripping is ubiquitous in the Coma cluster out to the virial radius.

\section*{Acknowledgements}

The authors thank Toby Brown, Ryan Chown, and Jacqueline Wightman for carefully classifying the visual morphology of Coma galaxies.  IDR thanks the Ontario Graduate Scholarship program for funding, LCP thanks the Natural Science and Engineering Council of Canada for funding.   This work was made possible thanks to a large number of publicly available software packages, including: \citep{astropy2013}, CMasher (https://cmasher.readthedocs.io), linmix (https://linmix.readthedocs.io/en/latest/index.html), Matplotlib \citep{hunter2007}, NumPy \citep{vanderwel2008}, scikit-learn \citep{scikit-learn}, SciPy \citep{jones2001}, statmorph \citep{rodriguez-gomez2019}, Stiff \citep{bertin2012}, Topcat
\citep{taylor2005}.
\par
The authors thank Dr. Stephen Gwyn for generating MegaCam image stacks for the central Coma field.  This work makes use of observations obtained with MegaPrime/MegaCam, a joint project of CFHT and CEA/DAPNIA, at the Canada-France-Hawaii Telescope (CFHT) which is operated by the National Research Council (NRC) of Canada, the Institut National des Sciences de l'Univers of the Centre National de la Recherche Scientifique of France, and the University of Hawaii.  This research used the facilities of the Canadian Astronomy Data Centre operated by the National Research Council of Canada with the support of the Canadian Space Agency.
\par
Funding for SDSS-III has been provided by the Alfred P. Sloan Foundation, the Participating Institutions, the National Science Foundation, and the U.S. Department of Energy Office of Science. The SDSS-III web site is http://www.sdss3.org/.
\par
SDSS-III is managed by the Astrophysical Research Consortium for the Participating Institutions of the SDSS-III Collaboration including the University of Arizona, the Brazilian Participation Group, Brookhaven National Laboratory, Carnegie Mellon University, University of Florida, the French Participation Group, the German Participation Group, Harvard University, the Instituto de Astrofisica de Canarias, the Michigan State/Notre Dame/JINA Participation Group, Johns Hopkins University, Lawrence Berkeley National Laboratory, Max Planck Institute for Astrophysics, Max Planck Institute for Extraterrestrial Physics, New Mexico State University, New York University, Ohio State University, Pennsylvania State University, University of Portsmouth, Princeton University, the Spanish Participation Group, University of Tokyo, University of Utah, Vanderbilt University, University of Virginia, University of Washington, and Yale University.




\bibliographystyle{mnras}
\bibliography{ref} 



\appendix

\section{ram pressure candidates}
\label{sec:image_appendix}

\begin{figure*}
    \centering
    \includegraphics[width=\textwidth]{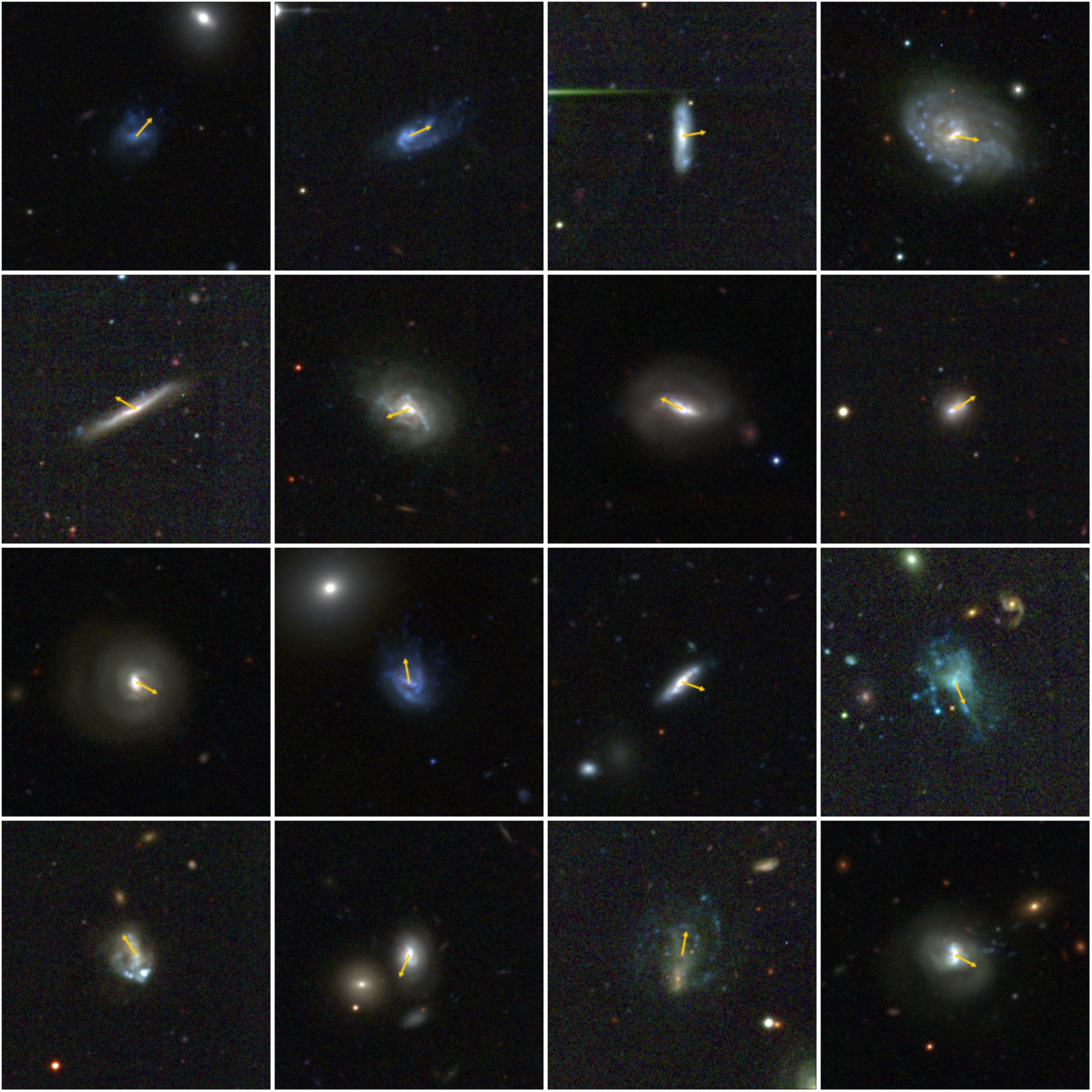}
    \caption{CFHT $ugi$ images for the sample of stripping candidates.  All cutout images have physical dimensions of $40 \times 40\,\mathrm{kpc}$.  The arrows in each thumbnail mark the estimated orientation of observed stripping features (see section~\ref{sec:orientation}}
    \label{fig:mosaic1}
\end{figure*}

\begin{figure*}
    \centering
    \includegraphics[width=\textwidth]{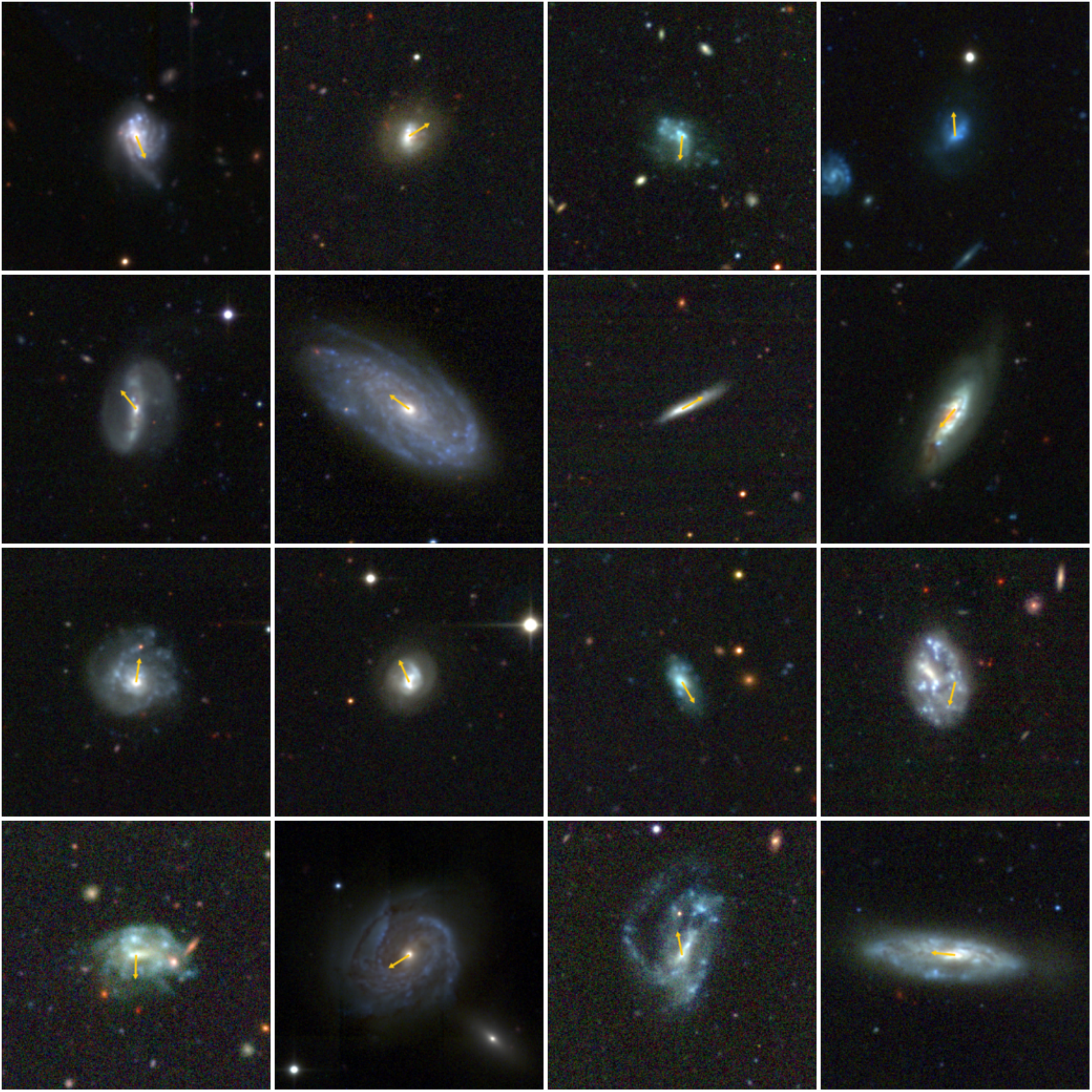}
    \caption{Continued from Fig.~\ref{fig:mosaic1}}
    \label{fig:mosaic2}
\end{figure*}

\begin{figure*}
    \centering
    \includegraphics[width=\textwidth]{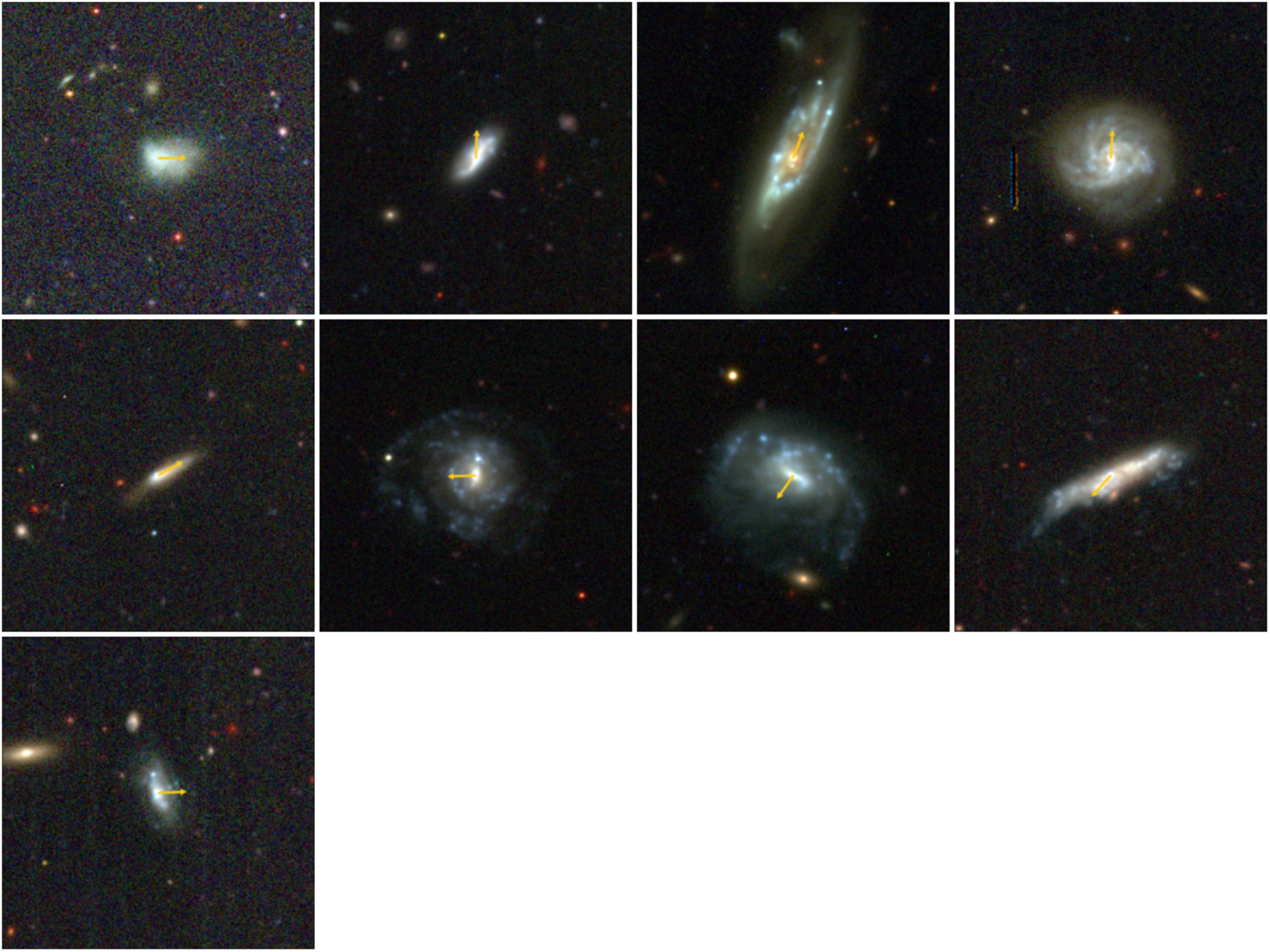}
    \caption{Continued from Fig.~\ref{fig:mosaic2}}
    \label{fig:mosaic3}
\end{figure*}

\begin{table*}
  \caption{Ram pressure candidates}
  \label{tab:sample}
  \centering
  \begin{threeparttable}
  \begin{tabular}{lccccc}
    \toprule
    Galaxy Name$^a$ & RA & Dec & $z$ & $\log M_\star$$^b$ & $\log \mathrm{SFR}$$^b$ \\[1.5pt]
    & deg & deg & & $\mathrm{M_\odot}$ & $\mathrm{M_\odot\,yr^{-1}}$ \\
    \midrule
    GMP 4629 & 194.4593 & 28.1704 & 0.0231 & 8.59 & -0.801 \\
    GMP 4570 & 194.4867 & 27.9918 & 0.0152 & 8.11 & -1.019 \\              
    GMP 5382 & 194.1197 & 29.1371 & 0.0316 & 9.33 & -0.475 \\            
    GMP 5422 & 194.1191 & 27.2913 & 0.0251 & 10.01 & -0.122 \\            
    GMP 2625 & 195.1300 & 28.9505 & 0.0233 & 9.18 & -0.974 \\             
    GMP 2599 & 195.1403 & 27.6377 & 0.0250 & 9.78 & 0.049 \\             
    GMP 1616 & 195.5328 & 27.6483 & 0.0230 & 10.24 & 0.404 \\            
    GMP 406 & 196.1616 & 28.9727 & 0.0253 & 9.06 & -0.664 \\               
    GMP 3779 & 194.7721 & 27.6444 & 0.0181 & 9.74 & 0.003 \\            
    GMP 3816 & 194.7586 & 28.1157 & 0.0314 & 10.16 & 0.560 \\             
    GMP 3618 & 194.8195 & 27.1061 & 0.0280 & 10.14 & 0.038 \\           
    GMP 5821 & 193.9346 & 28.7546 & 0.0275 & 8.91 & -0.448 \\            
    SDSS J130545.34+285216.8 & 196.4390 & 28.8713 & 0.0266 & 9.21 & -0.502 \\          
    GMP 2910 & 195.0381 & 27.8665 & 0.0177 & 9.27 & 0.080 \\             
    SDSS J130006.15+281507.8 & 195.0256 & 28.2522 & 0.0212 & 8.53 & -0.692 \\            
    GMP 4159 & 194.6472 & 27.2647 & 0.0245 & 9.78 & 0.242 \\          
    GMP 4135 & 194.6553 & 27.1766 & 0.0256 & 9.84 & 0.251 \\            
    GMP 4281 & 194.6064 & 28.1289 & 0.0274 & 9.72 & -0.300 \\                
    GMP 4236 & 194.6285 & 26.9949 & 0.0249 & 8.39 & -0.729 \\          
    GMP 3143 & 194.9552 & 26.9743 & 0.0235 & 8.89 & -0.820 \\            
    SDSS J130553.48+280644.7 & 196.4729 & 28.1124 &0.0246 & 10.09 & 0.170 \\          
    GMP 2544 & 195.1648 & 29.0194 & 0.0242 & 10.76 & -0.089 \\         
    GMP 4688 & 194.4321 & 29.0032 & 0.0231 & 8.51 & -0.881 \\              
    GMP 2559 & 195.1578 & 28.0580 & 0.0255 & 10.29 & 0.643 \\          
    GMP 3253 & 194.9172 & 28.6308 & 0.0178 & 9.37 & -0.315 \\              
    GMP 3271 & 194.9159 & 27.5765 & 0.0167 & 9.06 & -0.662 \\              
    GMP 6364 & 193.6752 & 27.6389 & 0.0287 & 8.53 & -0.856 \\                  
    GMP 1582 & 195.5473 & 28.1725 & 0.0299 & 8.70 & -0.187 \\           
    GMP 672 & 195.9880 & 26.7295 & 0.0220 & 8.70 & -0.796 \\
    GMP 2374 & 195.2336 & 27.7909 & 0.0266 & 11.32 & 0.501 \\             
    GMP 713 & 195.9768 & 28.3106 & 0.0268 & 8.74 & -0.250 \\             
    GMP 4437 & 194.5384 & 28.7086 & 0.0254 & 10.37 & 0.150 \\           
    GMP 4333 & 194.5822 & 28.0948 & 0.0239 & 8.52 & -0.733 \\             
    GMP 4463 & 194.5386 & 26.6641 & 0.0243 & 9.34 & -0.648 \\           
    GMP 4471 & 194.5233 & 28.2426 & 0.0240 & 10.88 & 1.083 \\
    GMP 2073 & 195.3545 & 28.6772 & 0.0292 & 10.35 & 0.321 \\             
    GMP 223 & 196.2776 & 28.6412 & 0.0182 & 8.58 & -0.718 \\           
    GMP 522 & 196.0945 & 28.8108 & 0.0265 & 9.86 & 0.159 \\            
    GMP 455 & 196.1106 & 27.3043 & 0.0184 & 9.26 & -0.274 \\             
    GMP 4106 & 194.6664 & 26.7595 & 0.0249 & 9.09 & -0.649 \\          
    GMP 597 & 196.0547 & 28.5425 & 0.0271 & 8.58 & -0.704 \\
    \bottomrule
  \end{tabular}
  \begin{tablenotes}
    \footnotesize
    \item \textsc{Notes.} $^a$ Galaxy ID from the \citet{godwin1983} catalog when applicable, otherwise from the SDSS; $^b$ Medium-deep GSWLC-2 catalogue \citep{salim2016,salim2018}
  \end{tablenotes}
  \end{threeparttable}
  
\end{table*}


\bsp	
\label{lastpage}
\end{document}